\documentclass[journal]{IEEEtran}
\usepackage[pdftex]{graphicx}

\usepackage[usenames]{color}
\usepackage{epsfig}
\usepackage{graphics}
\usepackage{amsmath}
\usepackage{amssymb}
\usepackage{multirow}
\usepackage{cite}
\usepackage{array}
\usepackage{pslatex} 
\usepackage{url}
\usepackage{booktabs}
\usepackage{caption}
\usepackage{lineno}
\usepackage{graphicx}  
\usepackage{setspace}
\usepackage{tikz}
\usepackage{hhline}
\usepackage{mathtools}
\usepackage[letterpaper]{geometry}
 \newcommand{\flogo}{\includegraphics[height=18pt]{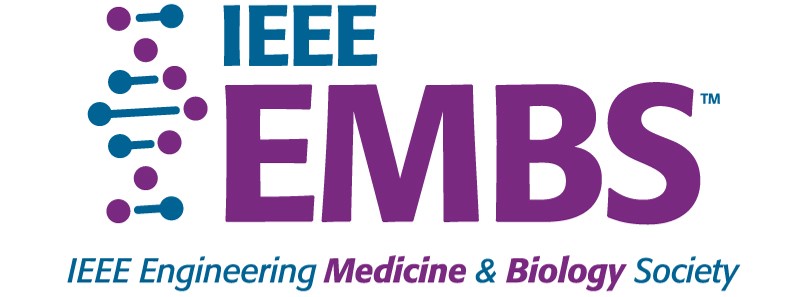}
 }
\ifCLASSINFOpdf
\else
\fi
\usepackage[english]{babel}
\usepackage[utf8]{inputenc}
\usepackage{fancyhdr}
\begin{document}
%
\title{\textcolor{violet}{Accurate and generalizable protein-ligand binding affinity prediction with geometric deep learning\vspace{0.25cm}}}

%
%
%

\author{Krinos Li, Xianglu Xiao, Zijun Zhong, Guang Yang*,~\IEEEmembership{Senior Member,~IEEE}
\thanks{
This work was supported in part by the ERC IMI (101005122), the H2020 (952172), the MRC (MC/PC/21013), the Royal Society (IEC/NSFC/211235), the NVIDIA Academic Hardware Grant Program, the SABER project supported by Boehringer Ingelheim Ltd, NIHR Imperial Biomedical Research Centre (RDA01), Wellcome Leap Dynamic Resilience, UKRI guarantee funding for Horizon Europe MSCA Postdoctoral Fellowships (EP/Z002206/1), UKRI MRC Research Grant, TFS Research Grants (MR/U506710/1), and the UKRI Future Leaders Fellowship (MR/V023799/1). 
	\par 
    K.L. \& X.X. Authors are with Bioengineering Department and Imperial-X, Imperial College London, London W12 7SL, UK. Z.Z Author is an independent researcher based in London W12 7LG, UK.
     *G. Y. Author is with a. Bioengineering Department and Imperial-X, Imperial College London, London W12 7SL, UK; 
    b. National Heart and Lung Institute, Imperial College London, London SW7 2AZ, UK; 
    c. Cardiovascular Research Centre, Royal Brompton Hospital, London SW3 6NP, UK; 
    d. School of Biomedical Engineering \& Imaging Sciences, King's College London, London WC2R 2LS, UK (correspondence e-mail: g.yang@imperial.ac.uk)
    }}

\maketitle\thispagestyle{fancy}

\begin{abstract}
\textit{Goal:} Protein-ligand binding complexes are ubiquitous and essential to life. Protein-ligand binding affinity prediction (PLA) quantifies the binding strength between ligands and proteins, providing crucial insights for discovering and designing potential candidate ligands. While recent advances have been made in predicting protein-ligand complex structures, existing algorithms for interaction and affinity prediction suffer from a sharp decline in performance when handling ligands bound with novel unseen proteins.
\textit{Methods:} We propose IPBind, a geometric deep learning-based computational method, enabling robust predictions by leveraging interatomic potential between complex's bound and unbound status. 
\textit{Results:} Experimental results on widely used binding affinity prediction benchmarks demonstrate the effectiveness and universality of IPBind. Meanwhile, it provids atom-level insights into prediction. 
\textit{Conclusions}: This work highlight the advantage of leveraging machine learning interatomic potential for predicting protein-ligand binding affinity. 
\end{abstract}


%
\IEEEpeerreviewmaketitle

\textbf{\textit{Impact Statement--}This study extends state-of-the-art deep learning algorithms to applications in protein-ligand binding affinity prediction. This study has implications for enhancing the generalization capability of protein-ligand interactions prediction methods by interatomic potential modeling.}\\
\\

\section{INTRODUCTION}

\begin{figure*}[t]
	\centering
	\includegraphics[width=\linewidth]{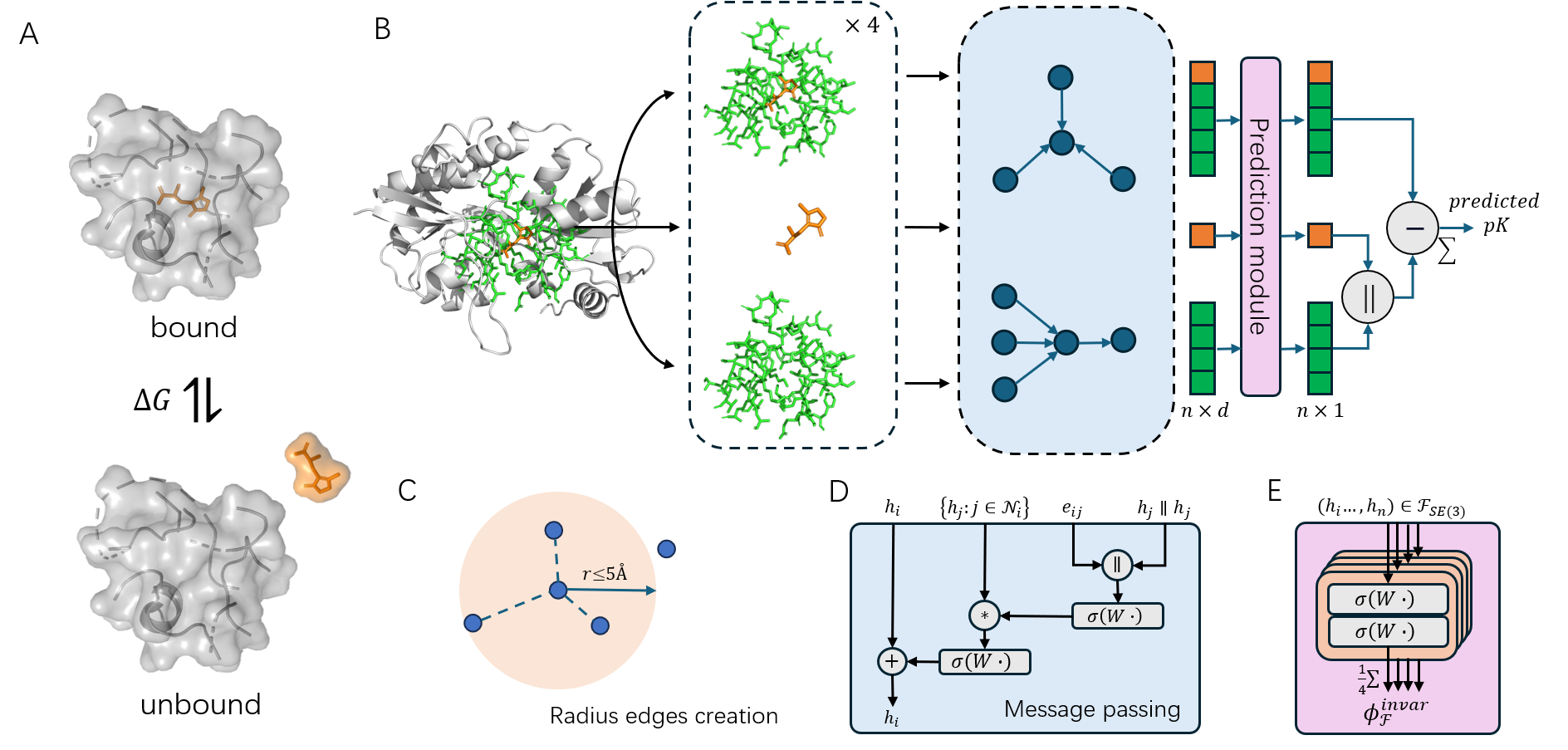}
	\caption{Overview of IPBind A. IPbind leverages the difference between bound and unbound status of protein-ligand to predict their binding affinity B. Workflow of IPBind. The initial protein-ligand complex were used to construct three atom-level graphs, after encoder module, atomic contribution of each graph will be summed up to predict the binding affinity between input protein and ligand. n: number of atoms in each molecule, d: number of representation dimensions. C. Radius edges creation D. Message passing layer. E. The prediction module of IPBind. }
	\label{fig:IPBind}
\end{figure*}
\IEEEPARstart{M}{odern} drug discovery is a protracted and costly endeavor, often spanning over a decade with expenditures exceeding billions of US dollars \cite{dimasi2016innovation}. A critical bottleneck lies in identifying high-affinity ligands that bind tightly to target proteins, as binding affinity—quantified by the energy difference between unbound and bound states—serves as a pivotal indicator of therapeutic potential. Higher binding affinity reflects stronger stabilization of the bound complex relative to the separated components \cite{hughes2011principles,guedes2018empirical}. While experimental techniques like isothermal titration calorimetry (ITC) and surface plasmon resonance (SPR) remain gold standards, their high cost and less efficiency hinder large-scale precessing \cite{du2016insights}.

The advent of structure-based computational approaches has reshaped early-stage drug discovery by enabling rapid in silico prediction. These methods leverage 3D structural insights to rationally design ligands that complement target binding sites \cite{francoeur2020three,sabe2021current}. Among them, deep learning (DL) has emerged as a transformative tool, and achieved breakthroughs in predicting protein-ligand affinities \cite{ozturk2018deepdta,nguyen2021graphdta,huang2020deeppurpose,somnath2021multi}. Unlike conventional methods requiring handcrafted feature engineering, DL models enables autonomously extracting hierarchical representations from structural data and capturing intricate patterns. For example, protein-ligand complex can be represented as voxels, and employs 3D convolutional neural networks (CNN) to learn the relationship between spatial interactions and physicochemical properties \cite{stepniewska2018development, jimenez2018k}. Modeling with graph neural networks (GNN) is a new research direction. A graph is a natural way to represent the structure of a molecule, in which each atom or residue can be set as a node and interactions between atoms can be treated as edges \cite{zhang2023ss,lu2022tankbind,lu2023highly,koh2024physicochemical}. 

Despite progress, critical challenges persist. First, the limited availability of high-quality experimental affinity data demands models with high data efficiency, which most existing DL-based PLA approaches rarely address \cite{durant2024future}. Second, poor cross-target generalization persists, particularly when deploying models across divergent protein families. This is reflected in model's performance drops when applied to novel receptor classes beyond training domains \cite{zhang2025artificial,wang2022learning,gao2023profsa}. Thirdly, many models overlook geometric invariance in their input representations. The lack of translational and rotational invariance in binding affinity predictions leads to inconsistent and unstable performance. Some existing works focus only on E(3) invariance, which fails to properly account for molecular chirality since E(3) network also treat reflection as invariance \cite{wu2024protein,yang2023geometric}.

To address these gaps, we propose IPBind, a efficient model that predicts protein-ligand binding affinity through \textit{Interatmoic Potential} modeling. Our key contributions are: 
(1): IPBind enables efficient training while achieving better generalization to unseen proteins. Additionally, it offers intuitive interpretability of predictions.
(2) IPBind is SE(3)-invariant, making it more aligned with real-world physics.
(3): Our model demonstrates robust performance across predicted structures as input, extending predictive capabilities beyond existing methods.
These characteristics highlight the potential of the proposed IPBind for real-world drug discovery pipelines.

\section{MATERIALS AND METHODS }
Our objective is to predict the binding affinity of protein-ligand complexes from their structures. This is a regression task where the labels represent continuous binding affinity values.

\subsection{Overview of IPBind}
We propose IPBind, a geometric deep learning model for predicting protein-ligand binding affinity. Our approach is inspired by the principle of machine-learning interatomic potentials, where the total energy of a molecular system is approximated as the sum of individual atomic contributions \cite{schutt2018schnet,fedik2022extending,batatia2022mace,batzner20223}. we derive the binding affinity by quantifying the sum of energy difference of each atom before and after binding. 
Figure \ref{fig:IPBind} illustrates the overall workflow of IPBind. 
The model takes protein-ligand complex as input. In order to make a balance between focusing on protein-ligand interactions and computational efficiency, we first cropped the protein pocket to concentrate on the most critical interaction sites, which was defined as maximum to the closest 50 residues to ligand's heavy atoms, and we preserve all heavy atoms of selected residues and ligand. The model consists of three main modules: preprocessing,  massage passing, and prediction.

To address the challenge of ensuring SE(3)-invariance for 3D molecules while avoiding the computational complexity, we employ frame averaging (FA) \cite{puny2021frame, duval2023hitchhiker}. FA introduces symmetry awareness to any architecture by averaging its outputs over a carefully selected subset of group elements, as frames. In the preprocessing module, three graphs were created for encoder models based on the input structure: the bound complex, the protein, and the ligand. We then create frames for the coordinates of each input. In the Encoder module, message passing is employed to capture interatomic interactions and iteratively update atomic representations.

The prediction module leverages encoder-generated atomic embeddings to predict the energy contribution of each atom in both its unbound and bound states. To ensure SE(3)-invariance, these predictions are averaged over the output of each frame. The final binding affinity is calculated as the sum of energy differences between the bound complex and unbound components.

\subsection{Preprocessing module}
\textbf{Frame averaging}
Since our goal is to predict binding affinity—-a scalar quantity—-we focus specifically on ensuring output invariance under SE(3) transformations. We adopt the Frame Averaging (FA) transformation to canonical plane to enable the output of geometric graph neural networks to be SE(3)-invariance \cite{puny2021frame}. 
For molecular systems with atom positions $X \in \mathbb{R}^{n \times 3}$, we define frames via principal component analysis (PCA):
\begin{enumerate}
    \item Compute centroid and covariance matrix:
    \begin{align}
        t &= \frac{1}{n}X^\top \mathbf{1} \in \mathbb{R}^3. \\
        \Sigma &= (X - \mathbf{1}t^\top)^\top(X - \mathbf{1}t^\top).
    \end{align}
    
    \item Solve eigenvalue problem $\Sigma u = \lambda u$ to obtain eigenvectors $u_1, u_2, u_3$ with $\lambda_1 > \lambda_2 > \lambda_3$
    
    \item Define 4 frames for achieving SE(3) invariant output using signed eigenbasis:
    \begin{equation}
        F_{\text{SE(3)}}(X) = \left\{ \left( [\alpha u_1, \beta u_2, u_1 \times u_2], \, t \right) \, \big| \, \alpha, \beta \in \{-1, 1\} \right\}.
    \end{equation}

\end{enumerate}

For above $(R, t) \in F_{\text{SE(3)}}(X)$, we averaging the result of above transformation to achieve SE(3)-invariant by:
\begin{align}
X_{\text{out}} = \frac{1}{|\mathcal{F_{\text{SE(3)}}}|} \sum_{(R,t) \in \mathcal{F_{\text{SE(3)}}}} \phi\left( (X - \mathbf{1}t)R, \, X \right),
\end{align}
that is, to translation and rotation invariance of the input coordinates. This adaptation preserves the theoretical guarantees from \cite{puny2021frame} while respecting biochemical reality in chirality. 

\textbf{Graph creation}
After frame averaging, we construct the geometric graphs based on  transformed position.
Three graphs incorporating bound complex, ligand and pocket atoms was built for input feature processing. We use 3D coordinates to construct radius edges at the atomic level with 5{\AA} cutoff (Figure \ref{fig:IPBind}.C), and treat each atom as a node.  

\subsection{Encoder module}

\textbf{Input embedding}
The initialization of node embeddings for each atom employs a multi-dimensional feature representation approach. Specifically, the atomic number is embedded into a 128-dimensional feature vector. For edge feature representation, we utilize the radial distribution function ($\text{RBF}(\cdot)$) to transform interatomic distances $d_{ij}$ and combine it with the relative position vectors $\tilde{\mathbf{r}}_{ij}$, ultimately embedding them into a 128-dimensional representation:

\begin{equation}
    \mathbf{e}_{ij} = \sigma\left(\text{MLP}\left(\text{RBF}(d_{ij}) \mathbin{\|} \tilde{\mathbf{r}}_{ij}\right)\right),
\end{equation}

where $\mathbin{\|}$ represents vector concatenation, and $\sigma(\cdot)$ is swish  function.

\textbf{Message passing}
We adopt the interatomic interaction layer proposed in FAENet \cite{duval2023faenet} to propagate atom-atom interactions defined through edges as in Figure \ref{fig:IPBind}.D.  During the message passing progress, only the node's representations are updated. Specifically, we learn an enhanced graph convolution filter:

\begin{equation}
    f^{(l)}_{ij} = \sigma\left(\text{MLP}\left(e_{ij} \mathbin{\|} \mathbf{h}^{(l)}_i \mathbin{\|} \mathbf{h}^{(l)}_j\right)\right),
\end{equation}

where $\mathbin{\|}$ represents vector concatenation. $\mathbf{h}^{(l)}_i \in \mathbb{R}^d$ denotes the feature vector of atom $i$ at layer $l$. The atomic features are updated through:

\begin{equation}
    \mathbf{h}^{(l+1)}_i = \mathbf{h}^{(l)}_i + \text{MLP}\left( \sum_{j \in \mathcal{N}_i} \mathbf{h}^{(l)}_j \odot f^{(l)}_{ij} \right),
\end{equation}

\noindent where $\odot$ denotes element-wise multiplication. $\mathcal{N}_i$ is the neighborhood of atom $i$.

\subsection{Prediction module}
A two-layer multilayer perceptron is used as the output head to predict the contribution of each atom in the prediction module. After that, we average the output of 4 frames for SE(3)-invariant prediction as in Figure \ref{fig:IPBind}.E. The predicted binding affinity is then calculated as the difference between the sum of atomic potentials from the unbound protein and ligand graphs and the sum from the protein-ligand complex graph.

\subsection{Model Training and optimization}
Given the regression nature of our task, we adopt the Balanced MSE loss \cite{ren2022balanced}, which explicitly accounts for label distribution skewness:

\begin{equation}
\begin{split}
    \mathcal{L}_{\mathrm{BME}} &= -\log p_{\text{train}}(\boldsymbol{y}|\boldsymbol{x};\boldsymbol{\theta}) \\
    &= -\log \left[ \frac{p_{\text{bal}}(\boldsymbol{y}|\boldsymbol{x};\boldsymbol{\theta}) \cdot p_{\text{train}}(\boldsymbol{y})}
        {\int_{\mathcal{Y}} p_{\text{bal}}(\boldsymbol{y}'|\boldsymbol{x};\boldsymbol{\theta}) \cdot p_{\text{train}}(\boldsymbol{y}') d\boldsymbol{y}'} \right] \\
    &\cong -\log \mathcal{N}(\boldsymbol{y}; \boldsymbol{y}_{\text{pred}}, \sigma^2_{\text{noise}}\mathbf{I}) \\
    &\quad + \log \int_{\mathcal{Y}} \mathcal{N}(\boldsymbol{y}'; \boldsymbol{y}_{\text{pred}}, \sigma^2_{\text{noise}}\mathbf{I}) \cdot p_{\text{train}}(\boldsymbol{y}') d\boldsymbol{y}'
\end{split}
\end{equation}

\noindent The noise variance \( \sigma^2_{\text{noise}} \) is implemented as a learnable parameter, jointly optimized with model \( \theta \).

We also incorporated an approximate NDCG loss \cite{valizadegan2009learning} to learn the relative binding strength of different protein-ligand complexes during training. Additionally, we applied exponential amplification to enhance its effectiveness:
\begin{equation}
  \mathcal{L}_{\mathrm{rank}} = \exp\left(\alpha(1 - \text{NDCG})\right) - 1 - \alpha(1 - \text{NDCG}).
\end{equation}
The total loss is defined as follows:
\begin{equation}
  \mathcal{L}_{\mathrm{Total}} = \mathcal{L}_{\mathrm{BME}} + \mathcal{L}_{\mathrm{rank}}
\end{equation}
We used AdamW as optimizer, and use a per-batch linear warm-up and cosine annealing strategy in training. The first two epochs serve as warm-up epochs, during which the learning rate increases linearly with the batch index from $1 \times 10^{-6}$ to 0.01. We set number of message passing layers as 4 for each independent run.

\begin{table*}[h]
\centering
\begin{minipage}{0.35\textwidth}
\centering
\small
\begin{tabular}{lll}
\toprule
Model & RMSE↓ & Pearson↑   \\
\midrule
Pafnucy & 1.450±0.047 & 0.769±0.019\\
SIGN & 1.295±0.010 & 0.807±0.004 \\

OnionNet & 1.399±0.076 & 0.770±0.027 \\

PotentialNet & 1.503±0.033 & 0.772±0.007 \\
GNN-DTI &  1.384±0.013 & 0.779±0.008\\
IGN & 1.269±0.030 & 0.821±0.013 \\ OnionNet-2 & \textbf{1.164} & \textbf{0.864} \\
SchNet & 1.390±0.023 & 0.787±0.016\\
EGNN & 1.289±0.021 & 0.816±0.011 \\
MetalProGNet & 1.309±0.043 & 0.802±0.013 \\ GraphDTA & 1.568±0.036 & 0.698±0.016 \\
GIGN & 1.190±0.017 & 0.840±0.007\\
SS-GNN & 1.165±0.011 & 0.846±0.004  \\ Tankbind & 1.402±0.021 & 0.773±0.008 \\
EHIGN & \underline{1.150±0.022} & \underline{0.854±0.004} \\ PSICHIC & 1.336±0.031 & 0.792±0.011 \\\hline
IPBind & 1.173±0.016 & 0.849±0.042  \\
\hline
\end{tabular}
\caption{Performance at CASF16 benchmark. The best one is bolded, the second best is underlined. Baselines are adopted from \cite{wang2021onionnet,yang2024interaction,koh2024physicochemical}}
\label{tab:casf}

\end{minipage}\hfill
\begin{minipage}{0.6\textwidth}
\centering
\small
\begin{tabular}{lcc|cc}
\hline
\multirow{2}{*}{Model} & \multicolumn{2}{c|}{Sequence Identity 30\%} & \multicolumn{2}{c}{Sequence Identity 60\%} \\
 & RMSE↓ & Pearson↑ & RMSE↓ & Pearson↑ \\
\hline
DeepDTA & 1.866±0.080 & 0.472±0.022 & 1.762±0.261 & 0.666±0.012 \\
B\&B & 1.985±0.006 & 0.165±0.006 & 1.891±0.004 & 0.249±0.006 \\
TAPE & 1.890±0.035 & 0.338±0.044 & 1.633±0.016 & 0.568±0.033 \\
ProtTrans & 1.544±0.015 & 0.438±0.053 & 1.641±0.016 & 0.595±0.014 \\
MaSIF & 1.484±0.018 & 0.467±0.020 & 1.426±0.017 & 0.709±0.008 \\
IEConv & 1.554±0.016 & 0.414±0.053 & 1.473±0.024 & 0.667±0.011 \\
Holoprot & 1.464±0.006 & 0.509±0.002 & 1.365±0.038 & 0.749±0.014 \\
ProtNet & 1.463±0.001 & 0.551±0.005 & 1.343±0.025 & 0.765±0.009 \\
EGNN-PLM & 1.403±0.010 & 0.565±0.020 & 1.559±0.020 & 0.644±0.020 \\
Uni-Mol & 1.520±0.030 & 0.558±0.000 & 1.619±0.040 & 0.645±0.020 \\
ProFSA & 1.377±0.010 & 0.628±0.010 & 1.377±0.010 & 0.764±0.000 \\
OnionNet-2 & 1.695±0.004 & 0.534±0.000 & 1.802±0.022 & 0.705±0.001 \\
PSICHIC & 1.561±0.011 & 0.478±0.001 & 1.290±0.004 & 0.763±0.001 \\
Tankbind & 2.041±0.049 & 0.305±0.003 & 1.462±0.007 & 0.741±0.000 \\
GIGN & 1.494±0.004 & 0.586±0.001 & \textbf{1.071±0.002} & 0.757±0.001 \\
EHIGN & \textbf{1.312±0.001} & \underline{0.612±0.000} & 1.274±0.002 & \underline{0.774±0.000}\\ \hline
IPBind & \underline{1.335±0.004} & \textbf{0.732±0.000} & \underline{1.119±0.000} & \textbf{0.843±0.000} \\
\hline
\end{tabular}
\caption{Performance comparison at LBA30 and LBA60. The best one is bolded, the second best is underlined. We ran OnionNet-2, PSICHIC, Tankbind, GIGN and EHIGN using their reported settings, and the rest of baselines are adopted from \cite{gao2023profsa}} 
\label{tab:atom3d}
\end{minipage}
\end{table*}
\begin{figure*}[h]
	\centering
	\includegraphics[width=0.9\linewidth]{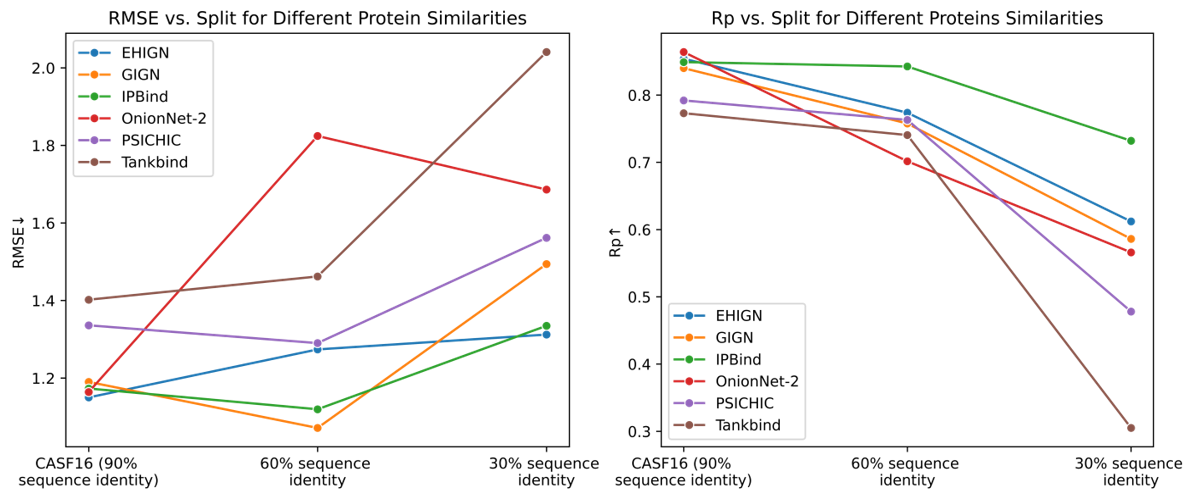}
	\caption{Performance comparison between model performance under different level of test protein sequence identity.}
	\label{fig:casfatom}
\end{figure*}

\section{RESULTS}
Our experiments are mainly focus on binding affinity prediction. Since the relative ranking of binding strengths is crucial in drug discovery, we primarily report Pearson correlation between labels and predicted values. We also use Root-Mean-Square-Error (RMSE) to evaluate the accuracy. 

\subsection{Dataset}
The proposed model was trained and evaluated on the refined set of PDBbind2020. We employed diverse benchmarks to evaluate IPBind. CASF2016 \cite{su2018comparative} is a commonly used benchmark for assessing binding affinity prediction performance. After removing 285 test samples, we used 5050 for training and randomly selected 500 for validation across five independent runs.

As CASF2016 shares high protein similarity with training set, which may lead to over-optimistic results \cite{kanakala2023latent,yang2024interaction}, we further adopted the LBA60\&LBA30 from Atom3D \cite{townshend2020atom3d}, restricting protein sequence identity to 60\% and harder 30\%, respectively, to better simulate real-world scenarios. A lower sequence identity indicates a lower likelihood of protein similarity between the training and test sets. The training/validation/test splits are 3563/448/452 for LBA60 and 3507/446/490 for LBA30. All experiments were repeated five times with different random seeds following prior work \cite{somnath2021multi,gao2023profsa,wang2022learning}.

\begin{figure*}[h]
	\centering
	\includegraphics[width=\linewidth]{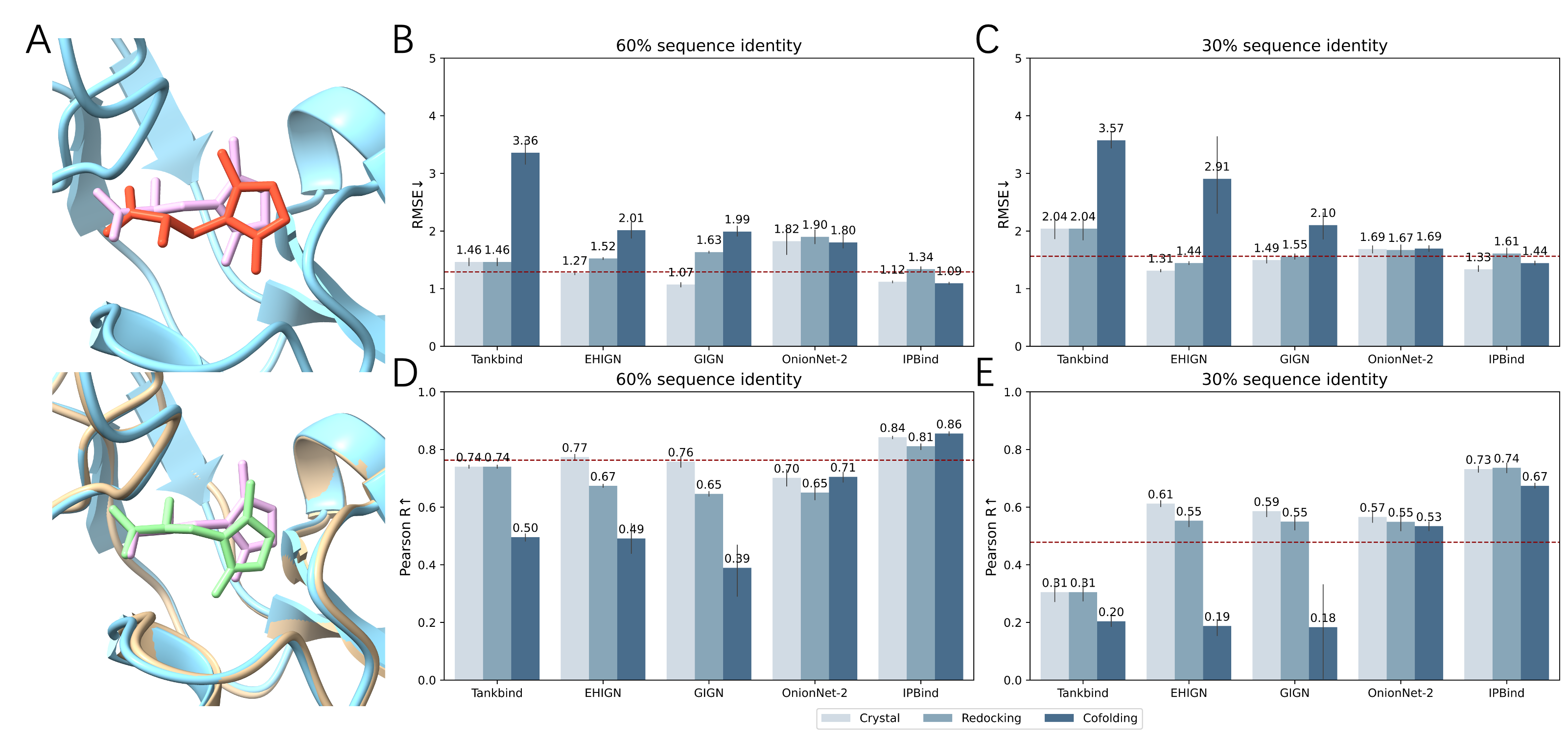}
	\caption{Model performance unbder different structure input. A. Difference between crystal structure vs redocking and colding structures in particular binding state with conformation PDB ID: 3DP4 (cyan: crystal protein, pink: crystal ligand, red: redocking ligand, beige: cofolding protein, green: cofolding ligand). B. \& C.  Model performance by different structure input in RMSE for LBA60 and 30. D. \& E. Model performance by different structure input in Pearson R for LBA60 and 30. Red horizontal line indicated the performance of PSICHIC, as 1.561, 1.290, 0.478, 0.763 in B., C., D., E., respectively.}
	\label{fig:dock}
\end{figure*}
\subsection{Results on CASF16}
We compared IPBind with wide arrange of previous studies \cite{li2021structure,stepniewska2018development,koh2024physicochemical,yang2023geometric,yang2024interaction,wang2021onionnet} at CASF16 in Table \ref{tab:casf}. The results show that IPBind performs excellently in RMSE and Pearson correlation coefficient. Specifically, IPBind outperforms other GNN-based methods such as GIGN and SS-GNN while maintaining comparable performance to the current state-of-the-art EHIGN. Notably, the moderate standard deviation in Pearson performance indicates consistent performance across diverse protein-ligand complexes.
\subsection{Results on dissimilar protein scenarios with Atom3D}
The advantage of IPBind becomes more pronounced when evaluated on the challenging Atom3D LBA splits designed for rigorous generalization testing. Table \ref{tab:atom3d} shows the results of the LBA30 and LBA60 benchmark tests of Atom3D.
For protein sequence similarity of train-test are reduced to 60\% identity, our model achieves an unprecedented Pearson correlation of 0.843, surpassing all baselines by $\geq$8.9\%. 
For the hardest low-sequence-similarity split at 30\% identity threshold, IPBind establishes new state-of-the-art performance with a Pearson correlation of 0.732, a 19.6\% relative improvement over previous best methods. The RMSE metrics confirm these advancements, corresponding to 4.1\% and 12.2\% error reductions from LBA60 to LBA30 versus EHIGN respectively. Meanwhile, compare with GIGN which build for E(3) invariance, our SE(3) method achieve 25\% better in Pearson and 11.9\% better in RMSE. These results further indicate the importance of considering physical rationality into modeling.

We further provide a comparison of the results from above three benchmarks in figure \ref{fig:casfatom}, with our IPBind and five baselines tested in all of these three benchmarks: PSICHIC, Tankbind, OnionNet-2, GIGN and EHIGN. We observed that as the test protein sequence identity decreases from 90\% (CASF16) to 30\%, the performance trends of various models become evident. This trend is consistent with the findings of previous work, suggesting that generalizing to proteins dissimilar to the training set remains a major challenge for binding affinity prediction models \cite{zhang2025artificial,gao2023profsa}. In terms of RMSE, IPBind demonstrates a relatively stable and moderate increasing pattern. Regarding the Pearson correlation metric, IPBind maintains a relatively high value across different sequence identities. This graphical evidence, in line with the tabular results from the previous benchmarks, further validates that that the IPBind model can still maintain advanced prediction performance in difficult scenarios.

\subsection{IPBind is robust to input structures}
Due to considerations of efficiency and cost, using crystal structures as model inputs is often impractical in real-world drug discovery, and predicted structures are typically employed instead. To further test the robustness of different models under practical conditions, we conducted additional tests on the more challenging LBA60 and LBA30 splits, considering two scenarios of predicted structures as inputs. The first is the lock-and-key case, where protein structure is considered a rigid body before and after binding \cite{holyoak2013molecular}. We redocked the protein's crystal structure with ligand molecules in this case
\cite{eberhardt2021autodock}. An other is the induced-fit scenario, the protein is regarded as flexible before and after binding with ligand and may differ from the known crystal structure. We illuminated the difference between these structures in figure \ref{fig:dock}.A. To simulate this situation, we using co-folding tools to generate structures from protein sequence and ligand SMILES\cite{chai2024chai,abramson2024accurate}.

We compared the performance of IPBind with that of PSICHIC, Tankbind, OnionNet-2, GIGN, and EHIGN, as depicted in Figures \ref{fig:casfatom} B-E. Tankbind utilizes separate protein and ligand inputs and therefore remains unaffected by redocking structures. Additionally, PSICHIC is a SOTA sequence-based model, which implies that it is immune to variations in the input model structures. Consequently, it serves as a baseline for comparing the robustness of other models that rely on structural inputs, and we represent it with a red dashed line in the figures.

Our findings indicate that GIGN/EHIGN showed minor declines under redocking structures, underperforming PSICHIC at 60\% sequence identity but outperforming slightly at 30\%. However, when using co-folded protein structures, these methods and Tankbind suffered sharp performance drops, lagging far behind PSICHIC.  

In contrast, IPBind and OnionNet-2 maintained robustness, with IPBind uniquely matching or exceeding PSICHIC across conditions. Notably, IPBind led in Pearson correlation on both benchmarks, demonstrating consistent performance with imperfect input structures—a critical advantage for practical applications where crystal structures are rarely available. This robustness highlights IPBind's reliability in real-world drug discovery pipelines reliant on predicted bound structures.

\begin{figure}[h]
	\centering
	\includegraphics[width=1\linewidth]{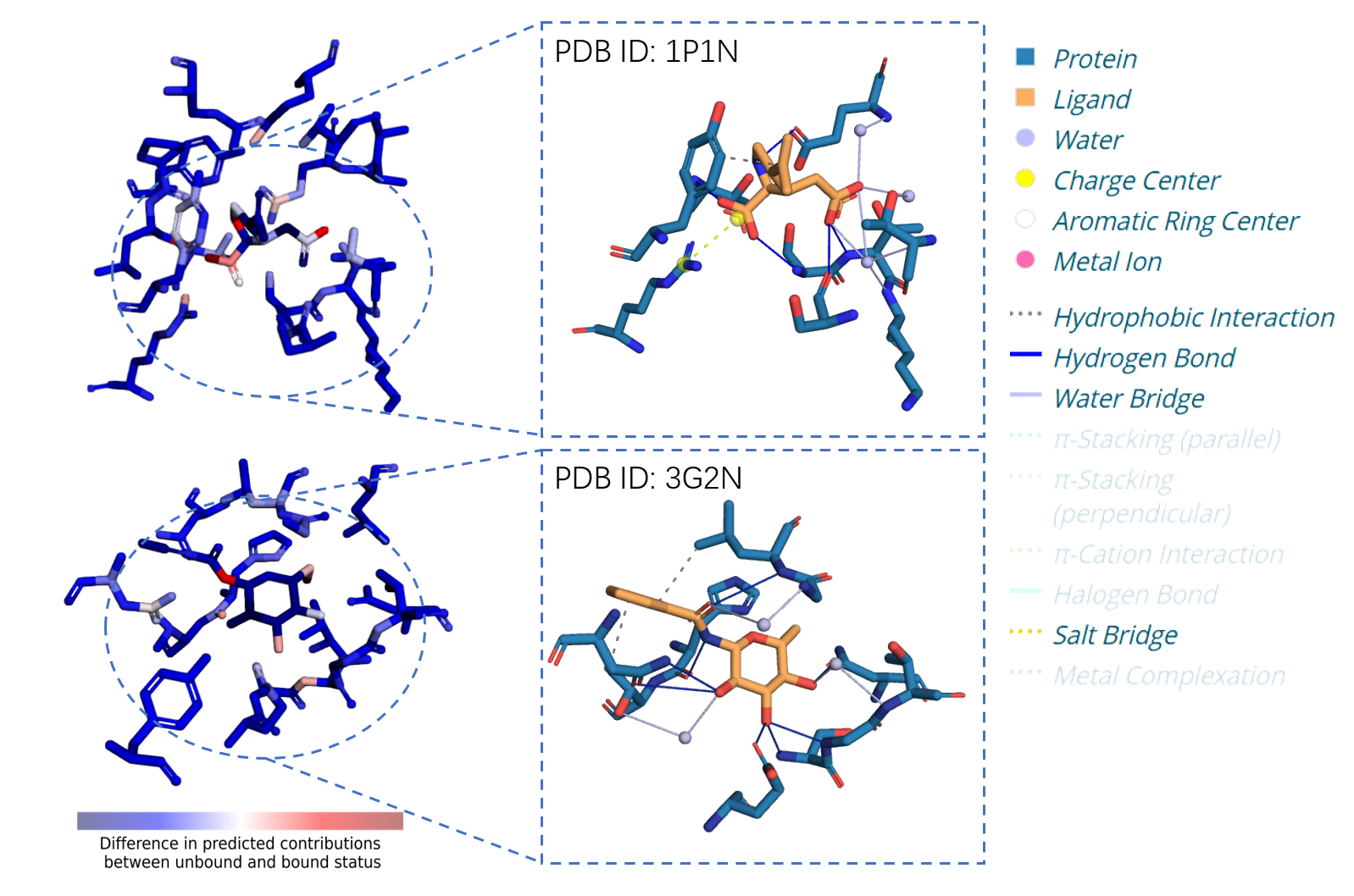}
	\caption{IPBind predictions vs PLIP interaction visualization.In the left-side, blue indicates a lower contribution of the atom to binding affinity, while red represents a higher contribution.}
	\label{fig:contribution}
\end{figure}

\begin{table*}[ht]
\centering

\begin{tabular}{lcc|cccc}
\hline
\multirow{2}{*}{Model} & {Group} & {Proposed}  & \multicolumn{2}{c}{Sequence Identity 30\%} & \multicolumn{2}{c}{Sequence Identity 60\%} \\&Symmetry&Framework
 & RMSE↓ & Pearson↑ & RMSE↓ & Pearson↑ \\
\hline
  IPBind&{SE(3)}&{Yes}         & 1.335±0.066   & 0.732±0.013         &   1.119±0.021     &   0.843±0.006    \\ 
\hline
   IPBind &{E(3)}&{Yes}         & 1.433±0.149         &   0.697±0.030   &         1.152±0.026&   0.829±0.005   \\ 
 IPBind&{N/A}&{Yes} &     1.607±0.087     &   0.676±0.009 &    1.266±0.027        & 0.802±0.008  \\ 
 IPBind$_{residue}$&{SE(3)}&{Yes} &   1.546±0.064     &   0.652±0.041 &    1.240±0.073        & 0.772±0.031  \\ 
 IPBind$_{residue}$&{SE(3)}&{No} &     1.625±0.067     &   0.624±0.049 &    1.420±0.087        & 0.694±0.055  \\ 
 IPBind$_{3 Networks}$&{SE(3)}&{Yes} &  1.776±0.037             & 0.411±0.007   & 1.832±0.016          & 0.521±0.009   \\ 

 IPBind&{SE(3)}&{No} &     2.116±0.017             & 0.302±0.003   & 1.896±0.011          & 0.419±0.008  \\ \hline
\end{tabular}
\caption{Results of ablation studies. }
\label{tab:method_comparison}
\end{table*}
\subsection{Output interpretability}
To interpret model predictions, we visualized atomic contributions from unbound/bound difference and compared them with Protein-Ligand Interaction Profiler (PLIP \cite{salentin2015plip}) analysis (Fig. \ref{fig:contribution}). For a well-predicted complex (1P1N, predicted 6.46 vs. true 6.8), IPBind correctly identified key binding-region atoms consistent with PLIP. In an underpredicted case (3G2N, predicted 5.41 vs. true 4.09), the model missed water-bridge-forming atoms, highlighting specific interaction types as potential improvement areas. These results further indicate that our framework design is simultaneously provide a valuable tool for understanding model predictions. When the predicted values deviate from the true values, by comparing the atom-contribution visualizations of IPBind and PLIP, researchers can easily spot which types of interactions or which specific atoms the model has overlooked or misjudged. 

\begin{figure}[h]
	\centering
	\includegraphics[width=0.9\linewidth]{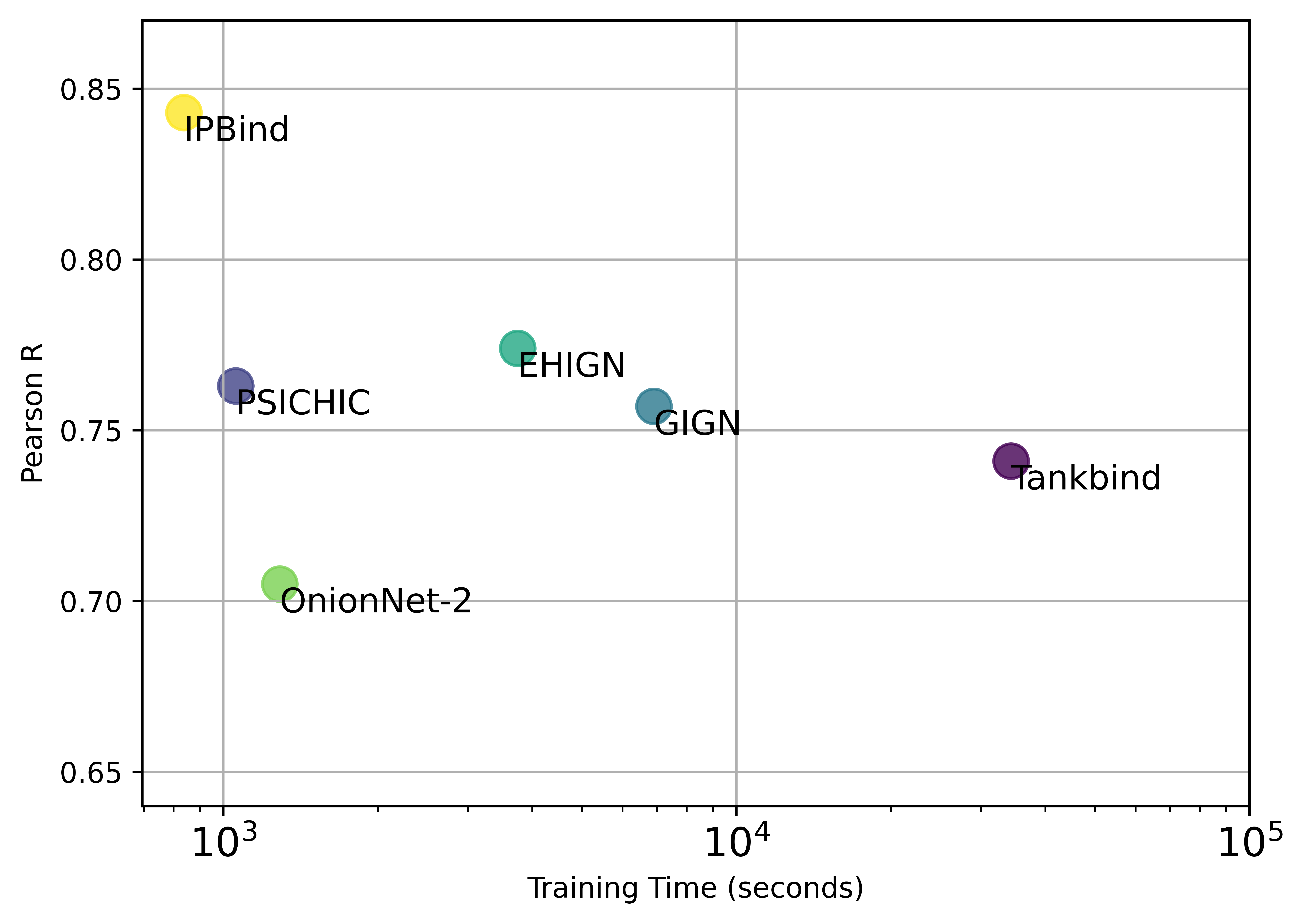}
	\caption{Performance vs average training time over 5 runs on LBA-60 benchmark}
	\label{fig:speed}
\end{figure}

\section{DISCUSSION}
IPBind’s design balances structural modeling and physical symmetry to improve binding affinity prediction. Ablation studies (Table \ref{tab:method_comparison}) confirm the critical role of SE(3) invariance: using E(3) or no symmetry group reduced Pearson R by up to 7.6\% at 30\% sequence identity, underscoring the importance of chirality-aware modeling. Modeling the difference between unbound and bound states of protein-ligand, outperforms vanilla approaches that only consider bound complex as input by 70\% in Pearson R. Using shared encoders (vs. IPBind$_{3 Networks}$ using separate networks for protein, ligand, and complex) also improving performance significantly. Meanwhile, we found replace protein into residue level graph IPBind$_{residue}$ could also harm the performance. Together these results validating the effectiveness of our unified architecture that leveraging molecular interactomic potentials for binding affinity prediction.

IPBind also offers superior training efficiency (Fig. \ref{fig:speed}), achieving state-of-the-art Pearson R on LBA60 within 14 minutes, and 8.9\% better than competitors with longer training times. Its minimal input requirements (atomic numbers and coordinates) eliminate costly preprocessing steps like charge calculation, making it highly deployable.

\section{CONCLUSION}
We proposed IPBind, a SE(3)-invariant method for protein-ligand binding affinity prediction. IPBind combines high prediction accuracy, structural robustness, interpretability, and computational efficiency, providing a promising tool for protein-ligand binding affinity prediction. Its performance across diverse benchmarks and realistic scenarios positions it as a valuable advancement for applications in drug discovery. 

\bibliography{references} 

\begin{thebibliography}{10}
\providecommand{\url}[1]{#1}
\csname url@samestyle\endcsname
\providecommand{\newblock}{\relax}
\providecommand{\bibinfo}[2]{#2}
\providecommand{\BIBentrySTDinterwordspacing}{\spaceskip=0pt\relax}
\providecommand{\BIBentryALTinterwordstretchfactor}{4}
\providecommand{\BIBentryALTinterwordspacing}{\spaceskip=\fontdimen2\font plus
\BIBentryALTinterwordstretchfactor\fontdimen3\font minus \fontdimen4\font\relax}
\providecommand{\BIBforeignlanguage}[2]{{%
\expandafter\ifx\csname l@#1\endcsname\relax
\typeout{** WARNING: IEEEtran.bst: No hyphenation pattern has been}%
\typeout{** loaded for the language `#1'. Using the pattern for}%
\typeout{** the default language instead.}%
\else
\language=\csname l@#1\endcsname
\fi
#2}}
\providecommand{\BIBdecl}{\relax}
\BIBdecl

\bibitem{dimasi2016innovation}
J.~A. DiMasi, H.~G. Grabowski, and R.~W. Hansen, ``Innovation in the pharmaceutical industry: new estimates of r\&d costs,'' \emph{Journal of health economics}, vol.~47, pp. 20--33, 2016.

\bibitem{hughes2011principles}
J.~P. Hughes, S.~Rees, S.~B. Kalindjian, and K.~L. Philpott, ``Principles of early drug discovery,'' \emph{British journal of pharmacology}, vol. 162, no.~6, pp. 1239--1249, 2011.

\bibitem{guedes2018empirical}
I.~A. Guedes, F.~S. Pereira, and L.~E. Dardenne, ``Empirical scoring functions for structure-based virtual screening: applications, critical aspects, and challenges,'' \emph{Frontiers in pharmacology}, vol.~9, p. 1089, 2018.

\bibitem{du2016insights}
X.~Du, Y.~Li, Y.-L. Xia, S.-M. Ai, J.~Liang, P.~Sang, X.-L. Ji, and S.-Q. Liu, ``Insights into protein--ligand interactions: mechanisms, models, and methods,'' \emph{International journal of molecular sciences}, vol.~17, no.~2, p. 144, 2016.

\bibitem{francoeur2020three}
P.~G. Francoeur, T.~Masuda, J.~Sunseri, A.~Jia, R.~B. Iovanisci, I.~Snyder, and D.~R. Koes, ``Three-dimensional convolutional neural networks and a cross-docked data set for structure-based drug design,'' \emph{Journal of chemical information and modeling}, vol.~60, no.~9, pp. 4200--4215, 2020.

\bibitem{sabe2021current}
V.~T. Sabe, T.~Ntombela, L.~A. Jhamba, G.~E. Maguire, T.~Govender, T.~Naicker, and H.~G. Kruger, ``Current trends in computer aided drug design and a highlight of drugs discovered via computational techniques: A review,'' \emph{European Journal of Medicinal Chemistry}, vol. 224, p. 113705, 2021.

\bibitem{ozturk2018deepdta}
H.~{\"O}zt{\"u}rk, A.~{\"O}zg{\"u}r, and E.~Ozkirimli, ``Deepdta: deep drug--target binding affinity prediction,'' \emph{Bioinformatics}, vol.~34, no.~17, pp. i821--i829, 2018.

\bibitem{nguyen2021graphdta}
T.~Nguyen, H.~Le, T.~P. Quinn, T.~Nguyen, T.~D. Le, and S.~Venkatesh, ``Graphdta: predicting drug--target binding affinity with graph neural networks,'' \emph{Bioinformatics}, vol.~37, no.~8, pp. 1140--1147, 2021.

\bibitem{huang2020deeppurpose}
K.~Huang, T.~Fu, L.~M. Glass, M.~Zitnik, C.~Xiao, and J.~Sun, ``Deeppurpose: a deep learning library for drug--target interaction prediction,'' \emph{Bioinformatics}, vol.~36, no. 22-23, pp. 5545--5547, 2020.

\bibitem{somnath2021multi}
V.~R. Somnath, C.~Bunne, and A.~Krause, ``Multi-scale representation learning on proteins,'' \emph{Advances in Neural Information Processing Systems}, vol.~34, pp. 25\,244--25\,255, 2021.

\bibitem{stepniewska2018development}
M.~M. Stepniewska-Dziubinska, P.~Zielenkiewicz, and P.~Siedlecki, ``Development and evaluation of a deep learning model for protein--ligand binding affinity prediction,'' \emph{Bioinformatics}, vol.~34, no.~21, pp. 3666--3674, 2018.

\bibitem{jimenez2018k}
J.~Jim{\'e}nez, M.~Skalic, G.~Martinez-Rosell, and G.~De~Fabritiis, ``K deep: protein--ligand absolute binding affinity prediction via 3d-convolutional neural networks,'' \emph{Journal of chemical information and modeling}, vol.~58, no.~2, pp. 287--296, 2018.

\bibitem{zhang2023ss}
S.~Zhang, Y.~Jin, T.~Liu, Q.~Wang, Z.~Zhang, S.~Zhao, and B.~Shan, ``Ss-gnn: a simple-structured graph neural network for affinity prediction,'' \emph{ACS omega}, vol.~8, no.~25, pp. 22\,496--22\,507, 2023.

\bibitem{lu2022tankbind}
W.~Lu, Q.~Wu, J.~Zhang, J.~Rao, C.~Li, and S.~Zheng, ``Tankbind: Trigonometry-aware neural networks for drug-protein binding structure prediction,'' \emph{Advances in neural information processing systems}, vol.~35, pp. 7236--7249, 2022.

\bibitem{lu2023highly}
S.~Lu, Z.~Gao, D.~He, L.~Zhang, and G.~Ke, ``Highly accurate quantum chemical property prediction with uni-mol+,'' \emph{arXiv preprint arXiv:2303.16982}, 2023.

\bibitem{koh2024physicochemical}
H.~Y. Koh, A.~T. Nguyen, S.~Pan, L.~T. May, and G.~I. Webb, ``Physicochemical graph neural network for learning protein--ligand interaction fingerprints from sequence data,'' \emph{Nature Machine Intelligence}, vol.~6, no.~6, pp. 673--687, 2024.

\bibitem{durant2024future}
G.~Durant, F.~Boyles, K.~Birchall, and C.~M. Deane, ``The future of machine learning for small-molecule drug discovery will be driven by data,'' \emph{Nature computational science}, pp. 1--9, 2024.

\bibitem{zhang2025artificial}
K.~Zhang, X.~Yang, Y.~Wang, Y.~Yu, N.~Huang, G.~Li, X.~Li, J.~C. Wu, and S.~Yang, ``Artificial intelligence in drug development,'' \emph{Nature Medicine}, pp. 1--15, 2025.

\bibitem{wang2022learning}
L.~Wang, H.~Liu, Y.~Liu, J.~Kurtin, and S.~Ji, ``Learning hierarchical protein representations via complete 3d graph networks,'' \emph{arXiv preprint arXiv:2207.12600}, 2022.

\bibitem{gao2023profsa}
B.~Gao, Y.~Jia, Y.~Mo, Y.~Ni, W.~Ma, Z.~Ma, and Y.~Lan, ``Profsa: Self-supervised pocket pretraining via protein fragment-surroundings alignment,'' \emph{arXiv preprint arXiv:2310.07229}, 2023.

\bibitem{wu2024protein}
K.~E. Wu, K.~K. Yang, R.~van~den Berg, S.~Alamdari, J.~Y. Zou, A.~X. Lu, and A.~P. Amini, ``Protein structure generation via folding diffusion,'' \emph{Nature communications}, vol.~15, no.~1, p. 1059, 2024.

\bibitem{yang2023geometric}
Z.~Yang, W.~Zhong, Q.~Lv, T.~Dong, and C.~Yu-Chian~Chen, ``Geometric interaction graph neural network for predicting protein--ligand binding affinities from 3d structures (gign),'' \emph{The journal of physical chemistry letters}, vol.~14, no.~8, pp. 2020--2033, 2023.

\bibitem{schutt2018schnet}
K.~T. Sch{\"u}tt, H.~E. Sauceda, P.-J. Kindermans, A.~Tkatchenko, and K.-R. M{\"u}ller, ``Schnet--a deep learning architecture for molecules and materials,'' \emph{The Journal of Chemical Physics}, vol. 148, no.~24, 2018.

\bibitem{fedik2022extending}
N.~Fedik, R.~Zubatyuk, M.~Kulichenko, N.~Lubbers, J.~S. Smith, B.~Nebgen, R.~Messerly, Y.~W. Li, A.~I. Boldyrev, K.~Barros \emph{et~al.}, ``Extending machine learning beyond interatomic potentials for predicting molecular properties,'' \emph{Nature Reviews Chemistry}, vol.~6, no.~9, pp. 653--672, 2022.

\bibitem{batatia2022mace}
I.~Batatia, D.~P. Kovacs, G.~Simm, C.~Ortner, and G.~Cs{\'a}nyi, ``Mace: Higher order equivariant message passing neural networks for fast and accurate force fields,'' \emph{Advances in neural information processing systems}, vol.~35, pp. 11\,423--11\,436, 2022.

\bibitem{batzner20223}
S.~Batzner, A.~Musaelian, L.~Sun, M.~Geiger, J.~P. Mailoa, M.~Kornbluth, N.~Molinari, T.~E. Smidt, and B.~Kozinsky, ``E (3)-equivariant graph neural networks for data-efficient and accurate interatomic potentials,'' \emph{Nature communications}, vol.~13, no.~1, p. 2453, 2022.

\bibitem{puny2021frame}
O.~Puny, M.~Atzmon, H.~Ben-Hamu, I.~Misra, A.~Grover, E.~J. Smith, and Y.~Lipman, ``Frame averaging for invariant and equivariant network design,'' \emph{arXiv preprint arXiv:2110.03336}, 2021.

\bibitem{duval2023hitchhiker}
A.~Duval, S.~V. Mathis, C.~K. Joshi, V.~Schmidt, S.~Miret, F.~D. Malliaros, T.~Cohen, P.~Li{\`o}, Y.~Bengio, and M.~Bronstein, ``A hitchhiker's guide to geometric gnns for 3d atomic systems,'' \emph{arXiv preprint arXiv:2312.07511}, 2023.

\bibitem{duval2023faenet}
A.~A. Duval, V.~Schmidt, A.~Hern{\'a}ndez-Garc{\i}a, S.~Miret, F.~D. Malliaros, Y.~Bengio, and D.~Rolnick, ``Faenet: Frame averaging equivariant gnn for materials modeling,'' in \emph{International Conference on Machine Learning}.\hskip 1em plus 0.5em minus 0.4em\relax PMLR, 2023, pp. 9013--9033.

\bibitem{ren2022balanced}
J.~Ren, M.~Zhang, C.~Yu, and Z.~Liu, ``Balanced mse for imbalanced visual regression,'' in \emph{Proceedings of the IEEE/CVF Conference on Computer Vision and Pattern Recognition}, 2022, pp. 7926--7935.

\bibitem{valizadegan2009learning}
H.~Valizadegan, R.~Jin, R.~Zhang, and J.~Mao, ``Learning to rank by optimizing ndcg measure,'' \emph{Advances in neural information processing systems}, vol.~22, 2009.

\bibitem{wang2021onionnet}
Z.~Wang, L.~Zheng, Y.~Liu, Y.~Qu, Y.-Q. Li, M.~Zhao, Y.~Mu, and W.~Li, ``Onionnet-2: a convolutional neural network model for predicting protein-ligand binding affinity based on residue-atom contacting shells,'' \emph{Frontiers in chemistry}, vol.~9, p. 753002, 2021.

\bibitem{yang2024interaction}
Z.~Yang, W.~Zhong, Q.~Lv, T.~Dong, G.~Chen, and C.~Y.-C. Chen, ``Interaction-based inductive bias in graph neural networks: enhancing protein-ligand binding affinity predictions from 3d structures,'' \emph{IEEE Transactions on Pattern Analysis and Machine Intelligence}, 2024.

\bibitem{su2018comparative}
M.~Su, Q.~Yang, Y.~Du, G.~Feng, Z.~Liu, Y.~Li, and R.~Wang, ``Comparative assessment of scoring functions: the casf-2016 update,'' \emph{Journal of chemical information and modeling}, vol.~59, no.~2, pp. 895--913, 2018.

\bibitem{kanakala2023latent}
G.~C. Kanakala, R.~Aggarwal, D.~Nayar, and U.~D. Priyakumar, ``Latent biases in machine learning models for predicting binding affinities using popular data sets,'' \emph{ACS omega}, vol.~8, no.~2, pp. 2389--2397, 2023.

\bibitem{townshend2020atom3d}
R.~J. Townshend, M.~V{\"o}gele, P.~Suriana, A.~Derry, A.~Powers, Y.~Laloudakis, S.~Balachandar, B.~Jing, B.~Anderson, S.~Eismann \emph{et~al.}, ``Atom3d: Tasks on molecules in three dimensions,'' \emph{arXiv preprint arXiv:2012.04035}, 2020.

\bibitem{li2021structure}
S.~Li, J.~Zhou, T.~Xu, L.~Huang, F.~Wang, H.~Xiong, W.~Huang, D.~Dou, and H.~Xiong, ``Structure-aware interactive graph neural networks for the prediction of protein-ligand binding affinity,'' in \emph{Proceedings of the 27th ACM SIGKDD conference on knowledge discovery \& data mining}, 2021, pp. 975--985.

\bibitem{holyoak2013molecular}
T.~Holyoak, ``Molecular recognition: lock-and-key, induced fit, and conformational selection,'' \emph{Encyclopedia of biophysics}, pp. 1584--1588, 2013.

\bibitem{eberhardt2021autodock}
J.~Eberhardt, D.~Santos-Martins, A.~F. Tillack, and S.~Forli, ``Autodock vina 1.2. 0: New docking methods, expanded force field, and python bindings,'' \emph{Journal of chemical information and modeling}, vol.~61, no.~8, pp. 3891--3898, 2021.

\bibitem{chai2024chai}
C.~D. team, J.~Boitreaud, J.~Dent, M.~McPartlon, J.~Meier, V.~Reis, A.~Rogozhonikov, and K.~Wu, ``Chai-1: Decoding the molecular interactions of life,'' \emph{BioRxiv}, pp. 2024--10, 2024.

\bibitem{abramson2024accurate}
J.~Abramson, J.~Adler, J.~Dunger, R.~Evans, T.~Green, A.~Pritzel, O.~Ronneberger, L.~Willmore, A.~J. Ballard, J.~Bambrick \emph{et~al.}, ``Accurate structure prediction of biomolecular interactions with alphafold 3,'' \emph{Nature}, vol. 630, no. 8016, pp. 493--500, 2024.

\bibitem{salentin2015plip}
S.~Salentin, S.~Schreiber, V.~J. Haupt, M.~F. Adasme, and M.~Schroeder, ``Plip: fully automated protein--ligand interaction profiler,'' \emph{Nucleic acids research}, vol.~43, no.~W1, pp. W443--W447, 2015.

\end{thebibliography}
\end{document}